\documentclass[aps,twocolumn,superscriptaddress,secnumarabic,balancelastpage,amsmath,amssymb,nofootinbib]{revtex4-1}
\usepackage{amssymb}
\usepackage{amsmath}
\usepackage{lgrind}        
\usepackage{color}         
\usepackage{graphics}      
\usepackage[pdftex]{graphicx}      
\usepackage{longtable}     
\usepackage{epsf}          
\usepackage{bm}            
\usepackage{asymptote}     
\usepackage{thumbpdf}
\usepackage{verbatim}
\usepackage{url}
\usepackage[colorlinks=true]{hyperref}  
\usepackage{enumitem}	
\newcommand{\ba}{\textbf{A}}

\begin{document}

\title{Perfect Single-Sided Radiation and Absorption without Mirrors}

\author{Hengyun Zhou}
\affiliation{Department of Physics, Massachusetts Institute of Technology, Cambridge, Massachusetts 02139, USA}
\author{Bo Zhen}
\email[Corresponding author: ]{bozhen@mit.edu}
\affiliation{Research Lab of Electronics, Massachusetts Institute of Technology, Cambridge, Massachusetts 02139, USA}
\affiliation{Physics Department and Solid State Institute, Technion, Haifa 32000, Israel}
\author{Chia Wei Hsu}
\affiliation{Department of Applied Physics, Yale University, New Haven, Connecticut 06520, USA}
\author{Owen D. Miller}
\affiliation{Department of Mathematics, Massachusetts Institute of Technology, Cambridge, Massachusetts 02139, USA}
\author{Steven G. Johnson}
\affiliation{Department of Mathematics, Massachusetts Institute of Technology, Cambridge, Massachusetts 02139, USA}
\author{John D. Joannopoulos}
\affiliation{Department of Physics, Massachusetts Institute of Technology, Cambridge, Massachusetts 02139, USA}
\affiliation{Research Lab of Electronics, Massachusetts Institute of Technology, Cambridge, Massachusetts 02139, USA}
\author{Marin Solja\v{c}i\'{c}}
\affiliation{Department of Physics, Massachusetts Institute of Technology, Cambridge, Massachusetts 02139, USA}
\affiliation{Research Lab of Electronics, Massachusetts Institute of Technology, Cambridge, Massachusetts 02139, USA}

\date{\today}

\begin{abstract}
Highly directional radiation from photonic structures is important for many applications, including high power photonic crystal surface emitting lasers, grating couplers, and light detection and ranging devices. 
However, previous dielectric, few-layer designs only achieved moderate asymmetry ratios, and a fundamental understanding of bounds on asymmetric radiation from arbitrary structures is still lacking. 
Here, we show that breaking the 180$^\circ$ rotational symmetry of the structure is crucial for achieving highly asymmetric radiation. 
We develop a general temporal coupled-mode theory formalism to derive bounds on the asymmetric decay rates to the top and bottom of a photonic crystal slab for a resonance with arbitrary in-plane wavevector. 
Guided by this formalism, we show that infinite asymmetry is still achievable even without the need of back-reflection mirrors, and we provide numerical examples of designs that achieve asymmetry ratios exceeding $10^4$. The emission direction can also be rapidly switched from top to bottom by tuning the wavevector or frequency. Furthermore, we show that with the addition of weak material absorption loss, such structures can be used to achieve perfect absorption with single-sided illumination, even for single-pass material absorption rates less than $0.5\%$ and without back-reflection mirrors. Our work provides new design principles for achieving highly directional radiation and perfect absorption in photonics.
\end{abstract}

\maketitle

\section{Introduction}

Due to their ease of fabrication and integration as well as their large area and high quality factor of resonances \citep{Lee2012}, photonic crystal slabs with one or two dimensional periodicity \cite{Johnson1999,Fan2002,Zhou2014} have been widely used in many applications, such as filters \cite{Suh2004a}, lasers \cite{Hirose2014}, and sensors \cite{Chow2004}. For more efficient utilization of light, it is often desirable to achieve highly directional out-of-plane coupling of light from photonic crystal slabs, in which light predominantly radiates to only one side of the slab. This would eliminate the need of a back-reflection mirror in high-power photonic crystal surface emitting lasers (PCSELs) \cite{Hirose2014}, where fabrication uncertainties in the laser wavelength and mirror-cavity distance currently make reliably achieving high slope efficiency difficult. This could also lead to increased efficiency of grating couplers for silicon photonics and light detection and ranging (LIDAR) devices. Previous designs of grating couplers have achieved a top-down asymmetry ratio (defined as the ratio of power going to the top and to the bottom) of up to 50:1 \cite{Taillaert2004, Roncone1993, Vermeulen2010, Subbaraman2015}, but they typically make use of a substrate reflector or involve multiple layers and grooves \cite{Fan2007}, which complicate fabrication and could be difficult to scale to larger areas if desired. Asymmetric out-of-plane emission from photonic crystal defect cavities of 4:1 has also been demonstrated \cite{Ota2015}, but all these works were guided primarily by numerical optimization. It is thus important to gain an understanding of the fundamental bounds on asymmetric radiation, and use such bounds as a guide to design stronger asymmetries.

Closely related to highly directional radiation is achieving perfect absorption of fields incident from a single side of a weakly-absorbing photonic structure. This can be viewed as the time-reversal partner of the single-sided radiation emission process. An increased absorption efficiency could be important for improving the performance of many devices, including modulators \cite{Liu2011}, photodetectors \cite{Xia2009}, solar cells \cite{Yu2010,Pospischil2014}. However, the single-pass absorption of a thin absorbing layer in air is at most 50$\%$ \cite{Hadley1947,Radi2013}. By combining electric and magnetic responses \textcolor{black}{or utilizing material anisotropy}, it is possible to design metamaterial perfect absorbers with near unity absorptance \cite{Radi2013,Landy2008,Watts2012,Baranov2015}, but such designs can be difficult to implement at optical frequencies. Recent work achieving perfect absorption in photonic crystal structures has either employed illumination from both sides and used the interference between the beams---analogous to a time-reversed laser---to achieve coherent perfect absorption \cite{Chong2010,Wan2011,Sun2014}, or employed a back-reflection mirror and critical coupling to resonances to approach perfect absorption \cite{Tischler2006,Lin2011,Piper2014,Liu2014,Zhu2015,Piper2016,Sturmberg2016}. Alternatively, specific surface textures can be designed to enhance light trapping and subsequent absorption\cite{Yu2010,Ganapati2014,Oskooi2014}, but a back-reflection mirror is still required to keep the photons inside the absorbing layer. In general, however, two-sided illumination can often be challenging to implement in realistic systems, while backing mirrors are often either lossy (e.g. metallic mirrors) or require additional fabrication efforts (e.g. distributed Bragg reflectors). Therefore, the possibility of achieving perfect absorption of fields incident from a single side, without the aid of backing mirrors, is highly attractive and could open up many engineering possibilities. \textcolor{black}{One recent approach achieving this is to utilize accidental degeneracies of critically coupled modes with opposite symmetries \cite{Piper2014a}, but such an approach requires aligning the frequencies and quality factors of multiple resonances.} Here, we first design structures with highly directional radiation, and then consider the time-reversal scenario at critical coupling to realize devices with high absorption efficiency. Moreover, away from the strongly-absorbing resonance frequency, light can be mostly transmitted through the designed devices, which could have important applications in multi-junction solar cells.

Previous work \cite{Wang2013} has used a temporal coupled-mode theory (TCMT) formalism \cite{Suh2004,Fan2003} with a single resonance and two coupling ports (one on each side of the slab) to examine bounds of asymmetric radiation from photonic crystal slabs. There, they reached the conclusion that the asymmetry ratio is bounded by $(1+r)/(1-r)$, where $r$ is the background amplitude reflection coefficient. For index contrasts found in realistic materials and at optical frequencies, this bound limits the proportion of radiation going to one side of the photonic crystal slab to around $90\%$, even for the high index contrast between silicon and air. For a smaller index contrast, this will be even more significantly different from perfect directional radiation. We find, however, that in more general scenarios, the bounds in Ref. \cite{Wang2013} can be greatly surpassed.

In a periodic photonic structure, the natural choice of mode basis is the momentum-conserving Bloch-wave basis \cite{Joannopoulos2011}. As shown in Fig.~\ref{fig:tcmt}(a), for general incident directions (nonzero in-plane momentum $\vec{k}_\parallel$) in asymmetric structures, the time-reversal operation relates the resonance at $\vec{k}_\parallel$ to the resonance at $-\vec{k}_\parallel$, suggesting that a two-resonance, four-port model is required to impose time-reversal constraints in the more general case. \textcolor{black}{Moreover, reciprocity automatically ensures that the two resonances share identical frequencies, eliminating the need for exquisite degenerate-frequency alignment to achieve multi-resonant responses.} Only when the system under consideration possesses certain symmetries---either in the structure \cite{Hsu2013a} ($C_2^z$, i.e. $180^\circ$ rotation around the out-of-plane axis) or in the incident field (normal incidence)---can we use the simplified model \cite{Wang2013} with only a single independent port on each side of the slab.

In this article we show that the general two-resonance, four-port TCMT formalism, widely applicable to periodic structures with arbitrary geometry or in-plane momentum, enables bounds with significantly higher (sometimes even infinitely high) radiation asymmetry for realistic materials when $C_2^z$ symmetry of the structure is broken. As an example, we apply this formalism to inversion-symmetric ($P$-symmetric) structures without $C_2^z$ symmetry. Through numerical examples, we show that a top-down asymmetry ratio exceeding $10^4$ can be achieved by tuning the resonance frequency to coincide with the perfectly transmitting frequency on the Fabry-Perot background. The emission direction can also be rapidly switched from top to bottom by tuning the $\vec{k}_\parallel$ vector or frequency. These results provide important design principles for PCSELs, grating couplers, LIDARs, and many other applications that could benefit from directional emission and rapid tuning. In addition, we derive analytical expressions for the transmission spectrum and discuss features such as full transmission or reflection. We then show that such highly asymmetric coupling to the two sides of the photonic crystal slab can also be employed to achieve perfect absorption of light incident from one side of the slab, without the need of back-reflection mirrors as in previous designs.

\section{Temporal Coupled-Mode Theory Formalism}

\begin{figure}[h!tb]
\centering
\fbox{\includegraphics[width=\linewidth]{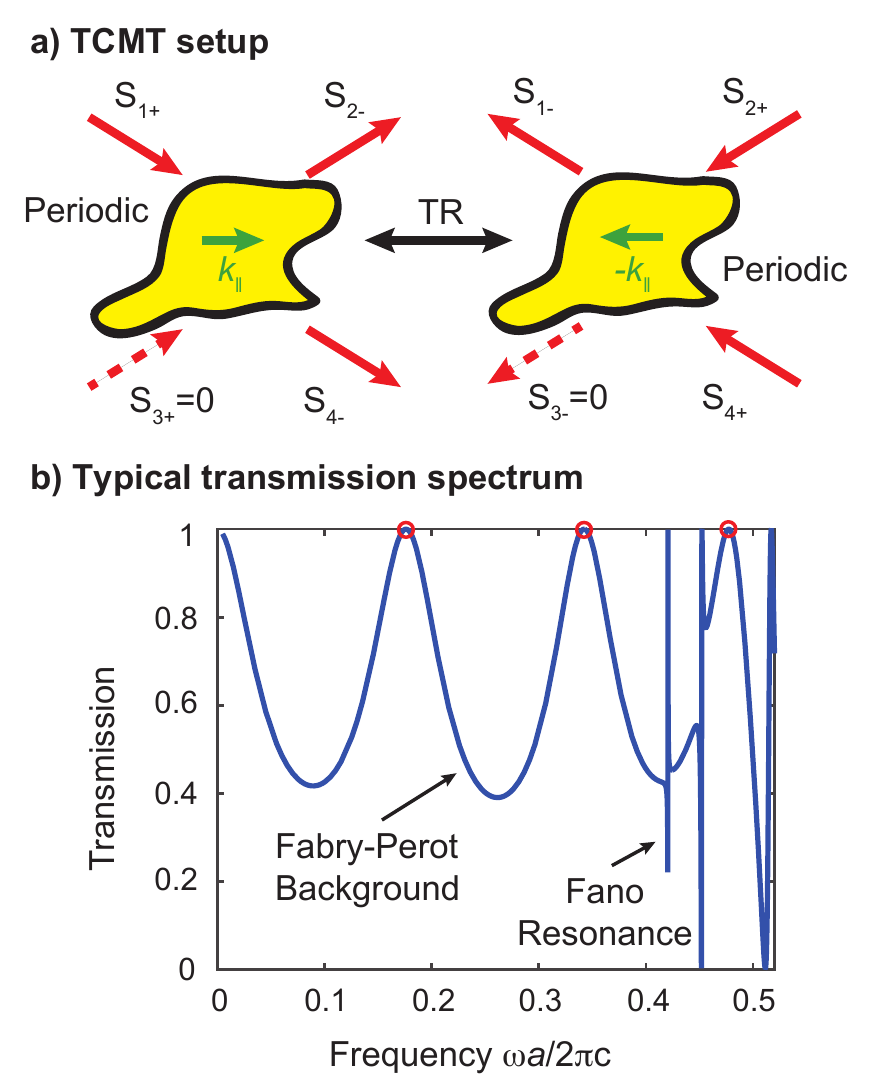}}
\caption{Temporal coupled-mode theory (TCMT) setup and transmission spectrum. (a) Schematic of our TCMT setup with four ports and two resonances related by the time reversal operation. This general setup is valid for structures with arbitrary shapes and incident angles as long as the assumption of four ports and two resonances is correct. (b) Typical transmission spectrum of an inversion-symmetric, $C_2^z$-symmetry-broken structure, with the Fano resonances exhibiting full transmission at certain frequencies as predicted by our TCMT formalism. Strong asymmetry is achieved when the Fano resonance is aligned with the frequencies where the background reaches full transmission (red circles).}
\label{fig:tcmt}
\end{figure}

We start by considering arbitrary photonic crystal slab structures embedded in a uniform medium (identical substrate and superstrate). We assume weak coupling, linearity, energy conservation and time-reversal symmetry in the system, and we consider frequencies below the diffraction limit so that higher-order diffractions are not present. A plane wave with in-plane momentum $\vec{k}_\parallel=(k_x,k_y)$ incident from the top [port 1, see Fig.~\ref{fig:tcmt}(a)] will only couple to resonances and outgoing waves with the same $\vec{k}_\parallel$ (conservation of Bloch momentum). We shall consider the typical case where there is a single resonance at $\vec{k}_\parallel$ near the frequencies of interest, with the transmission spectrum consisting of a Fabry-Perot background and sharp resonant features, as in Fig.~\ref{fig:tcmt}(b). To describe time-reversal symmetry constraints for general geometries and incident angles, we need to include the resonance at $-\vec{k}_\parallel$ in our description as well, resulting in a two-resonance, four-port model. Although conservation of Bloch momentum means that each input port only excites either the $\vec{k}_\parallel$ or $-\vec{k}_\parallel$ resonance, the two resonances still influence one another indirectly via the time-reversal symmetry constraints on the coupling matrices between resonances and ports ($K$ and $D$ in Eq. (\ref{eq:TCMT1})), as described below. Writing down expressions consistent with momentum conservation and time-reversal symmetry, we obtain the TCMT equations
\begin{align}
\frac{d\ba}{dt}&=\left(j\omega-\frac{1}{\tau}-\frac{1}{\tau_{nr}}\right)\ba+K^{\rm T}|s_+\rangle,\; |s_-\rangle=C|s_+\rangle+D\ba,\label{eq:TCMT1}\\
C&=e^{j\phi}\begin{pmatrix}
0 & r & 0 & jt\\r & 0 & jt & 0\\0 & jt & 0 & r\\jt & 0 & r & 0\\
\end{pmatrix},\; D=K\sigma_x=\begin{pmatrix}
0 & d_1\\d_2 & 0\\0 & d_3\\d_4 & 0\\
\end{pmatrix},\label{eq:TCMTcoeff}
\end{align}
where $\ba=(A_1,A_2)^{\rm T}$ are the amplitudes of the two resonances (at $\vec{k}_\parallel$ and $-\vec{k}_\parallel$ respectively), $\omega$ is the resonance frequency shared by both resonances, $\tau$ is the radiative $e^{-1}$-decay lifetime ($\omega$, $\tau$ are identical for the two resonances due to reciprocity), $\tau_{nr}$ is the $e^{-1}$-decay lifetime for nonradiative processes such as absorption, and $|s_{+}\rangle=(s_{1+},s_{2+},s_{3+},s_{4+})^{\rm T}$, $|s_{-}\rangle=(s_{1-},s_{2-},s_{3-},s_{4-})^{\rm T}$ are the amplitudes of the incoming and outgoing waves. $C$ is the scattering matrix for the direct (non-resonant) transmission and reflection through the slab (namely, the Fabry-Perot background). Energy conservation and reciprocity constrain $C$ to be unitary and symmetric. For identical substrates and superstrates, $C$ takes the form in Eq.~(\ref{eq:TCMTcoeff}), where $t$ and $r$ are real numbers satisfying $r^2+t^2=1$ that characterize the Fabry-Perot background, and the phase $\phi$ depends on the choice of reference plane position. $K$ and $D$ are the coupling matrices in and out of the resonances. Time reversal flips the two resonances, so instead of the usual relation $D=K$ \cite{Suh2004}, here we have $D=K\sigma_x$, where $\sigma_x$ is the $2\times 2$ $X$-Pauli matrix acting to flip the resonances. We note that an alternative (and equivalent) formalism is to adopt a basis in which the underlying modes are time-reversal invariant (for which the standard multimode treatment \cite{Suh2004} is adequate), by superimposing the resonance at $\vec{k}_\parallel$ with its time-reversal partner at $-\vec{k}_\parallel$. This is to be contrasted with a basis change for ports, as discussed in Ref. \cite{Ruan2012}. Detailed derivations of this and the following expressions are given in \href{link}{Supplement 1}. 

Energy conservation and time-reversal symmetry impose constraints on the coefficients. Energy conservation requires
\begin{align}\label{eq:econs}
|d_2|^2+|d_4|^2=|d_1|^2+|d_3|^2=\frac{2}{\tau},
\end{align}
while time-reversal symmetry gives the constraint $D=K\sigma_x$ and the two independent equations
\begin{align}\label{eq:fulld1}
e^{j\phi}(rd_2^*+jtd_4^*)+d_1&=0,\\
e^{j\phi}(jtd_2^*+rd_4^*)+d_3&=0.\label{eq:fulld2}
\end{align}
In the following, we shall fix the phase $\phi$ to be 0 by appropriately choosing the location of our reference plane. Eqs.~(\ref{eq:econs}, \ref{eq:fulld1}, \ref{eq:fulld2}) impose constraints on the values and phases of the couplings, and hence constrain the transmission spectrum and set bounds on the asymmetric coupling ratios.

From the preceding equations, we can derive an expression for the transmission spectrum \cite{Wang2013,Fan2003} that only depends on the frequencies and decay rates of the resonances and the transmission and reflection coefficients of the direct Fabry-Perot pathway.

The full scattering matrix including the direct pathway and resonance pathway \cite{Suh2004} is given by  Eq.~(\ref{eq:scatmat}) in \href{Link}{Supplement 1}. The power reflection and transmission coefficient for a wave incident from port 1 correspond to the amplitude squared of the (1,2), (1,4) element of the scattering matrix, given by
\begin{align}\label{eq:R}
R&=|S_{12}|^2=\left|e^{j\phi}r+\frac{d_1d_2}{j(\omega-\omega_0)+\frac{1}{\tau}+\frac{1}{\tau_{nr}}}\right|^2,\\\label{eq:T}
T&=|S_{14}|^2=\left|e^{j\phi}jt+\frac{d_1d_4}{j(\omega-\omega_0)+\frac{1}{\tau}+\frac{1}{\tau_{nr}}}\right|^2,
\end{align}
and the power reflection coefficient can be rewritten in the lossless limit $\tau_{nr}\rightarrow\infty$ as
\begin{align}\label{eq:nolossR}
R&=\frac{\left[r(\omega-\omega_0)\pm\sqrt{\frac{4}{\tau_1\tau_2}-\frac{r^2}{\tau^2}-\frac{2}{\tau\sigma}-\frac{1}{\sigma^2 r^2}}\right]^2+\left(\frac{1}{\sigma r}\right)^2}{(\omega-\omega_0)^2+\frac{1}{\tau^2}},
\end{align}
where we have written $\tau_i=2/|d_i|^2$, $1/\sigma=1/\tau_1-1/\tau_4$ to simplify the expression.

This expression provides general conditions for reaching full transmission or reflection with the Fano resonance. As shown in \href{link}{Supplement 1}, full transmission $R=0$ can only occur when the coupling rates satisfy $\tau_1=\tau_4$, $\tau_2=\tau_3$ ($P$-symmetric coupling), consistent with the transmission spectrum shown in Fig.~\ref{fig:tcmt}(b). Full reflection $R=1$ can only occur when the coupling rates satisfy $\tau_1=\tau_2$, $\tau_3=\tau_4$ ($C_2^z$ symmetric coupling), consistent with the results in Ref.~\cite{Wang2013}. Note that for structures that do not have $P$ or $C_2^z$ symmetry, it is still possible for the coupling rates for resonances to be $P$ or $C_2^z$ symmetric, leading to full transmission/reflection features in the frequency spectrum (see for example \href{link}{Supplement 1}, Fig. \ref{fig:multispec}).

\section{General Bounds on Asymmetric Coupling Rates}


We now derive bounds on the achievable asymmetry of coupling to the top and bottom based on Eqs.~(\ref{eq:fulld1}, \ref{eq:fulld2}) derived from time-reversal symmetry. Denote $|d_4/d_2|=a_r$, $|d_3/d_1|=a_\ell$, and define the asymmetric coupling ratios on the right ($\vec{k}_\parallel$) and left ($-\vec{k}_\parallel$) directions of the resonator as $a_r^2$ and $a_\ell^2$ (the ratio of the power going to bottom and top). By taking the ratio of Eqs.~(\ref{eq:fulld1}, \ref{eq:fulld2}), we find
\begin{align}
a_\ell^2=\left|\frac{jt+ra_re^{j\theta}}{r+jta_re^{j\theta}}\right|^2=\frac{t^2+r^2a_r^2+2tra_r\sin\theta}{r^2+t^2a_r^2-2tra_r\sin\theta},
\end{align}
where $\theta=\arg(d_2)-\arg(d_4)$ characterizes the phase difference between $d_2$ and $d_4$. This gives the bound
\begin{align}\label{eq:fullbound}
\left|\frac{t-ra_r}{r+ta_r}\right|\leq a_\ell\leq \left|\frac{t+ra_r}{r-ta_r}\right|.
\end{align}
\begin{figure}[htb]
\centering
\fbox{\includegraphics[width=\linewidth]{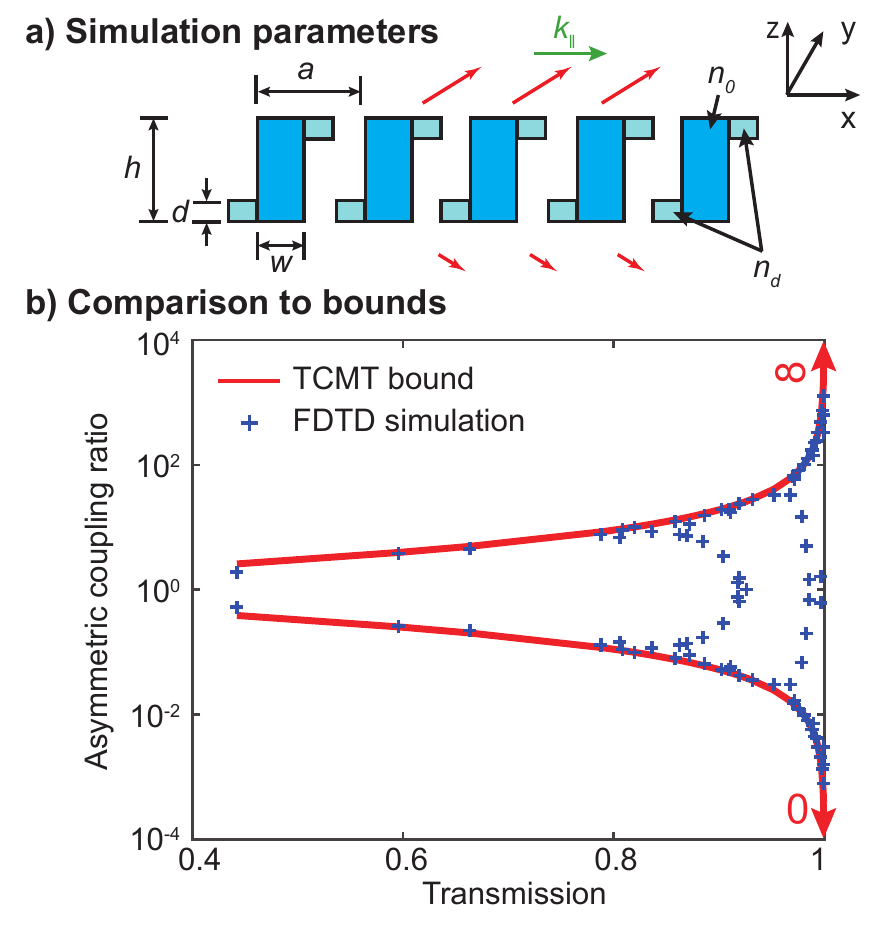}}
\caption{Simulated structures and verification of TCMT bounds. (a) The $P$-symmetric structure we use in our numerical examples and its structural parameters. $a$: periodicity of photonic crystal, $h$: height of central slab, $w$: width of central slab, $n_0$: refractive index of central slab, $d$: height of additional pieces on the sides (the width of the additional pieces is $(a-w)/2$), $n_d$: refractive index of additional pieces on the sides; (b) Numerical verification of TCMT bounds on asymmetric radiation for $P$-symmetric structures. Red lines indicate the bound from Eq.~(\ref{eq:tbound}). Each blue cross indicates simulation results of the asymmetry for a given structure, optimized over in-plane momentum. The transmissivity $t$ is fitted from the Fabry-Perot background, and the asymmetric coupling ratio is calculated from the Poynting flux in the top and bottom directions.
\label{fig:bound}}
\end{figure}
Therefore, the amount of achievable asymmetry to top and bottom on the left is bounded by that on the right, and vice versa. Note that in general the phase here can be tuned through a $2\pi$ cycle, so the bounds---even up to infinitely high asymmetry ratio---should be saturable for appropriate parameter choices. \textcolor{black}{The coefficients that enter the bounds are the transmission/reflection coefficients ($t,r$) of the direct process (Fabry-Perot background), as opposed to the total transmission/reflection including the resonant pathway.}

If the structure has $C_2^z$ symmetry, the two channels on the top and bottom will be constrained to have the same coupling rates, so $d_1=d_2$, $d_3=d_4$, $a_\ell=a_r$. Plugging this into Eq.~(\ref{eq:fullbound}), we find the same bound as in Ref.~\cite{Wang2013}:
$\frac{1-r}{1+r}\leq a_\ell^2=a_r^2\leq\frac{1+r}{1-r},$
which shows the consistency of our approach. In typical photonic crystal systems at optical frequencies, the index contrast between the slab and the background medium is limited to around 3, which constrains the interface reflection coefficient to be less than 0.5 for most incident angles. This results in the Fabry-Perot direct pathway reflection coefficient $r=\sqrt{1-t^2}$ being considerably smaller than 1, so strong asymmetry in the decay rates is difficult to achieve for $C_2^z$-symmetric structures without the use of an additional back-reflecting mirror.

The general bound Eq.~(\ref{eq:fullbound}) suggests, however, that much stronger asymmetry can be achieved if we break the $C_2^z$ symmetry of the system. A simple example is when the structure possesses inversion symmetry $P$ but breaks $C_2^z$ symmetry, as shown in Fig.~\ref{fig:bound}(a). In this case, the decay rates must satisfy $d_1=d_4$, $d_2=d_3$, $a_\ell=1/a_r$, and Eq.~(\ref{eq:fullbound}) becomes
\begin{align}\label{eq:tbound}
\frac{1-t}{1+t}\leq a_\ell^2=\frac{1}{a_r^2}\leq \frac{1+t}{1-t}.
\end{align}
For any index contrast, due to the up-down symmetry of the background material, the Fabry-Perot background will always have frequencies with full transmission, as exemplified by the red circles in Fig.~\ref{fig:tcmt}(b). Therefore, by tuning the resonance frequency to such points, the lower and upper bounds of Eq.~(\ref{eq:tbound}) approach $0$ and $+\infty$. Moreover, the bound can be saturated for appropriate choices of structural parameters and wavevectors, yielding arbitrarily high asymmetric decay rates of the photonic structure to top and bottom directions. \textcolor{black}{We note that similar design principles of breaking $C_2^z$ symmetry to achieve higher radiation asymmetry have also been realized in Ref. \cite{Wade2015} using the different design intuition of destructive interference.}


To verify these analytical results, we perform numerical simulations using the finite difference time domain (FDTD) method \cite{Taflove2013} with a freely available software package \cite{Oskooi2010}. We extract the coupling rate to top or bottom by monitoring the field amplitude at reference planes placed in the far-field, and determine the Fabry-Perot background transmissivity from plane wave excitation calculations. The results are given in Fig.~\ref{fig:bound}(b), where each data point in the figure (blue crosses) represents the maximal asymmetric coupling ratio searching over all $\vec{k}$ points in the Brillouin zone, for $P$-symmetric structural parameter choices (Fig.~\ref{fig:bound}(a)) with $h=1.5a$, $w=0.45a$, $n_0=1.45$, and varying $n_d$ and $d$ (see \href{Link}{Supplement 1} for more details). We can see that all data points obey the bound Eq.~(\ref{eq:tbound}) derived above (red solid lines). Moreover, this bound can be saturated for each value of the background transmission coefficient by appropriate optimization of the structural parameters and in-plane momentum. The blue crosses that do not saturate the bound are structures with very little perturbation from $C_2^z$ symmetry due to the choice of structural parameters.

\section{Examples of Highly Asymmetric Radiation}

In this section, we provide numerical examples of strong asymmetry that highlight two features of the extreme data points in Fig.~\ref{fig:bound} that are not obvious from the preceding data: it is possible to achieve high asymmetry even at the point of highest quality factor, and it is possible to achieve rapid tuning of the direction of asymmetry by slightly changing the frequency. As the form $(1+t)/(1-t)$ of the bound (for $P$-symmetric structures) suggests, strong asymmetry can be achieved when the resonance frequency coincides with locations of large transmissivity on the Fabry-Perot background, for any refraction index contrast.
\begin{figure}[htb]
\centering
\fbox{\includegraphics[width=\linewidth]{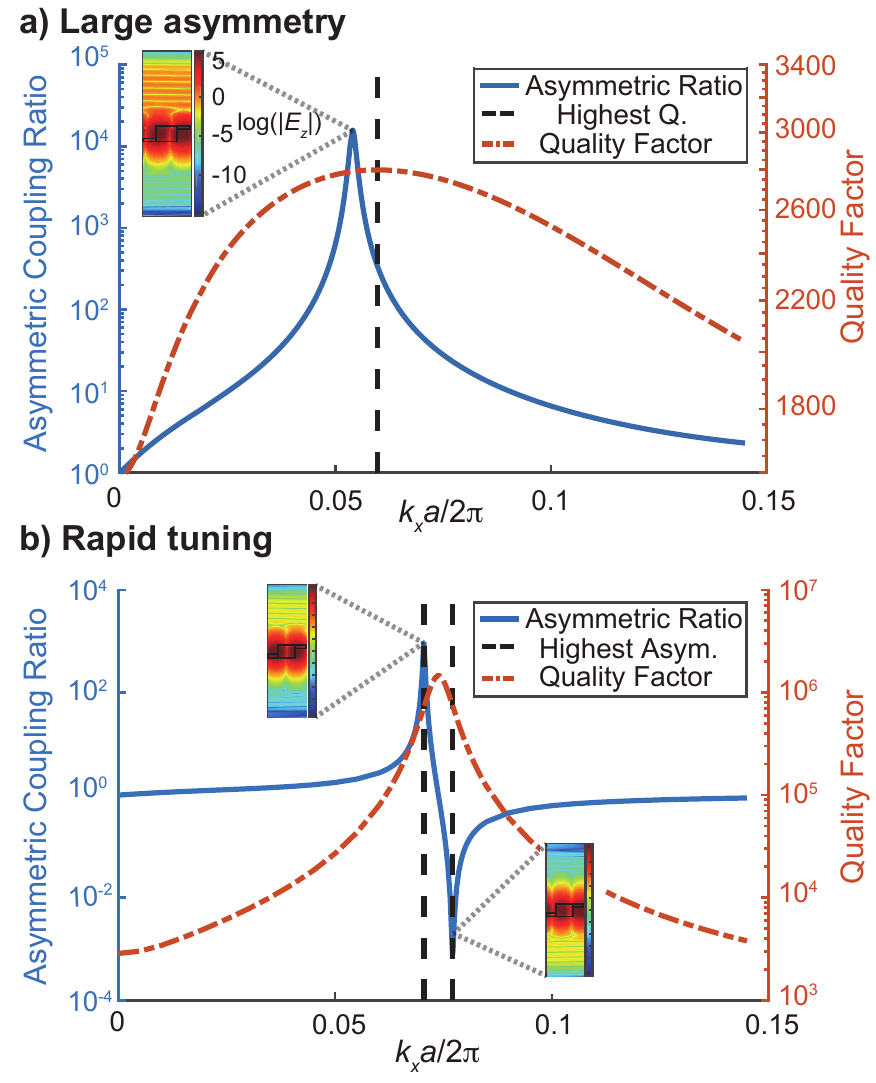}}
\caption{Examples of highly asymmetric radiation. (a) Plot of the asymmetry ratio and quality factor as a function of $k_x$, along the $k_y=0$ axis in momentum space. Strong asymmetric radiation occurs over a range of momenta, including the point of highest quality factor. \textcolor{black}{Inset: log scale plot of the $z$-component of the electric field amplitude at the highest asymmetry point;} (b) Similar plot for a different set of parameters, showing rapid switching of asymmetric direction by tuning the frequency or in-plane momentum.
\label{fig:maxasym}}
\end{figure}

We optimize over the structural parameters shown in Fig.~\ref{fig:bound}(a) to find examples of high asymmetry in coupling to the top and bottom. This example consists of the second transverse electric (TE) polarization band (nonzero $E_z, E_x, H_y$, classified by mirror-symmetry with respect to the $x-z$ plane) of a 1D photonic crystal with structural parameters $h=1.5a$, $w=0.45a$, $n_0=n_d=1.45$, $d=0.3a$, as defined in Fig.~\ref{fig:bound}(a). See \href{link}{Supplement 1} for a plot of the band structure. The resulting asymmetry ratio and quality factor ($Q$) as a function of the in-plane $k_x$, along with the radiation field distribution at maximal asymmetry, are shown in Fig.~\ref{fig:maxasym}(a). The resonance frequency lies very close to a point of full transmission on the Fabry-Perot background, and exhibits an asymmetry exceeding $10^4$ at the $\vec{k}_\parallel$ point of largest asymmetry as well as an asymmetry over $300$ at the point of highest quality factor. It may therefore be possible to produce a laser that preferentially emits to the top or bottom using the principles discussed above.

Another application of our results is the rapid steering of the direction of light emission by slight tuning of the frequency, which could be useful for LIDARs \cite{Yaacobi2014} or antennas. We design such a structure by perturbing a bound state in the continuum (BIC) \cite{Hsu2013a,Zhen2014,Hsu2013b,Hsu2016}. BICs are localized solutions embedded in the radiation continuum, where due to destructive interference of the amplitude for decay between outgoing wave channels, the quality factor of a resonance above the light line approaches infinity. In previous work \cite{Hsu2013a,Zhen2014,Hsu2013b}, the photonic structures were chosen to have both $P$ and $C_2^z$ symmetries. With a perturbation that breaks $C_2^z$ but preserves $P$, the peak quality factor will be finite but still very high \cite{Hsu2013b}. We expect that this symmetry breaking will also split the momenta where radiation towards the top and towards the bottom vanish, thereby creating strong asymmetry in the two directions, with the extrema separated only by a small $\vec{k}_\parallel$.

We choose $h=1.5a$, $w=0.45a$, $n_0=1.45$, $d=0.1a$, $n_d=1.1$, again examining the second TE band. The resulting asymmetry ratio and frequency are shown in Fig.~\ref{fig:maxasym}(b). The asymmetric coupling flips from mostly radiating to the top to mostly radiating to the bottom (by a factor of $10^4$) when $k_x$ is changed by as small as $0.05\times 2\pi c/a$ or equivalently, when the frequency is changed by $3\times 10^{-4}\times 2\pi c/a$. The radiative quality factor of these resonances are on the order of $10^6$, so these two bands will be well-separated in emission. One can thereby envision rapid tuning of the emission direction by changing the frequency of radiation slightly. Moreover, the high $Q$ of these resonances will enable long propagation lengths for collimated emission from large areas that is important for LIDAR applications, complementing the low-$Q$ designs of conventional grating couplers.

\begin{figure}[h!tb]
\centering
\fbox{\includegraphics[width=\linewidth]{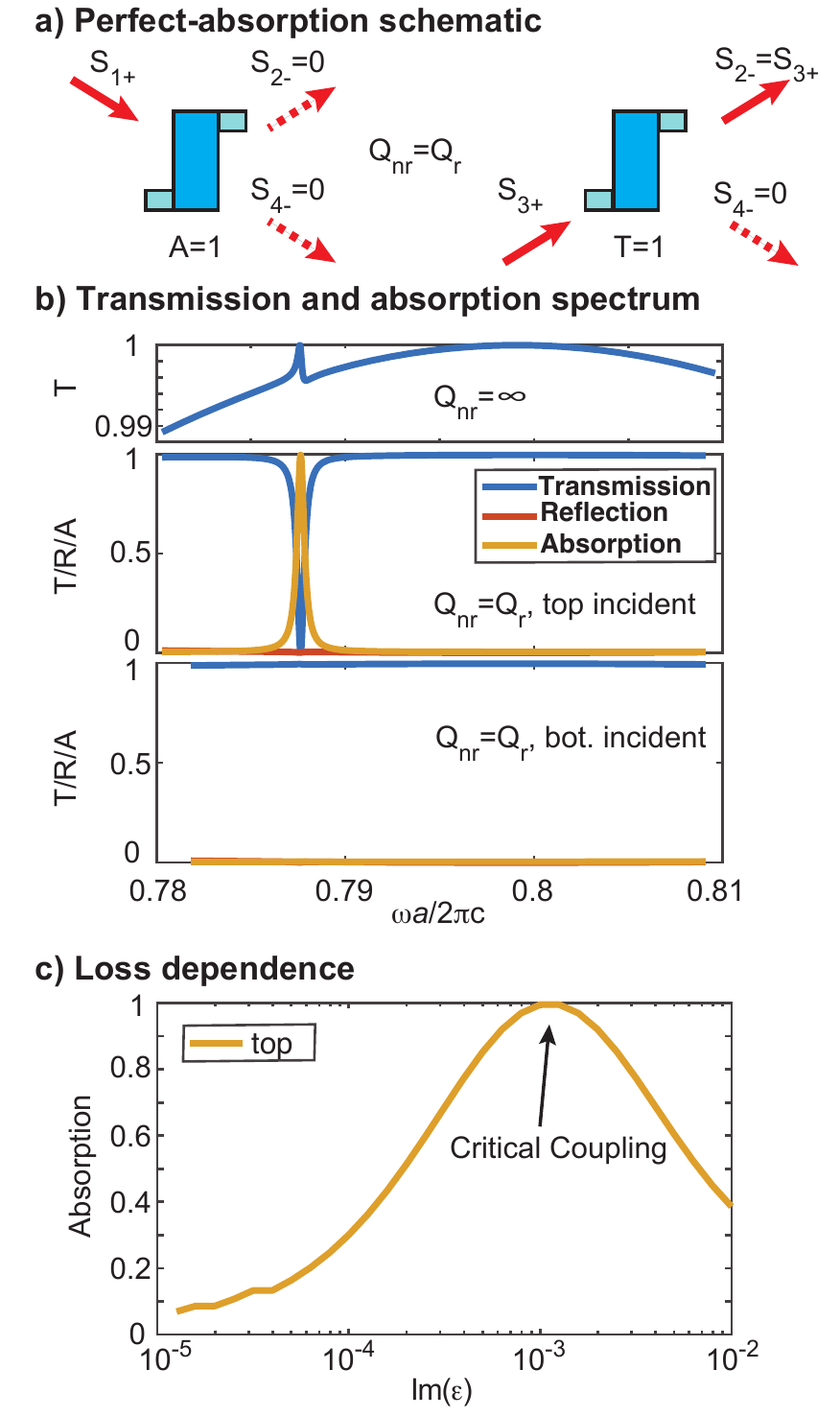}}
\caption{Perfect absorption with single-sided illumination and no backing mirror, for single-pass absorption less than $0.5\%$. (a) Schematic for perfect absorption at one incident angle and perfect transmission at the opposite incident angle. (b) Transmission, reflection and absorption spectra for no loss ($Q_{nr}=\infty$) and critical loss ($Q_{nr}=Q_r$), consistent with the theoretical results in (a). (c) Loss dependence of absorption, showing near-perfect absorption for critical coupling.
\label{fig:absorb}}
\end{figure}

\section{Perfect Absorption with Single-Sided Illumination and No Backing Mirrors}
We now discuss achieving perfect absorption in photonic crystal structures by combining the highly asymmetric coupling to different channels and matched radiative and non-radiative quality factors. Previous work on achieving perfect absorption has utilized metamaterial responses \cite{Radi2013,Landy2008,Watts2012}, interference between multiple incident directions \cite{Chong2010,Wan2011,Sun2014}, or a backing mirror to confine and trap light \cite{Tischler2006,Lin2011,Piper2014,Liu2014,Zhu2015,Piper2016,Sturmberg2016}. On the other hand, our results on achieving highly directional coupling suggest that it may be possible to achieve near-perfect absorption with single-sided illumination from the direction with strong coupling, only using dielectric structures and without the need of any backing mirrors. Intuitively, since the radiation coupling to one of the emission channels is strongly suppressed, there is only one direction to which the excited resonance can radiatively decay into. Appropriate tuning of the quality factor can then result in destructive interference towards this direction, thus achieving perfect absorption of incident waves.

To incorporate material loss in our description, we include a non-radiative decay channel, setting $\tau_{nr}$ to be finite in Eq.~(\ref{eq:TCMT1}). We assume that the loss rate is small, and that the direct transmission pathway is not affected by the loss \cite{Piper2014}. We start from a $P$-symmetric structure and incident direction where the coupling rates have a large asymmetry ratio. The parameters are chosen such that $t$ is close to 1, and the asymmetry ratio saturates the bound $|d_2|/|d_1|=\sqrt{(1+t)/(1-t)}$. As shown in \href{Link}{Supplement 1}, $P$-symmetry and the time-reversal constraints Eqs.~(\ref{eq:fulld1}, \ref{eq:fulld2}) imply that $d_1=d_4=\sqrt{(1-t)/\tau}\exp(-3\pi j/4)$, $d_2=d_3=\sqrt{(1+t)/\tau}\exp(-\pi j/4)$. Plugging this into the expressions for the power transmission and reflection coefficients, we find that on resonance $\omega=\omega_0$ and at critical coupling $\tau=\tau_{nr}$,
\begin{align}
R_{14}&=R_{23}=\frac{r^2}{4},\quad T_{14}=\left(\frac{1+t}{2}\right)^2,\quad T_{23}=\left(\frac{1-t}{2}\right)^2,\\
A_{14}&=1-R_{14}-T_{14}=\frac{1-t}{2},\quad A_{23}=\frac{1+t}{2}.
\end{align}
In the limit where $t\rightarrow 1$, $r\rightarrow 0$, it follows that light incident from port 1 or 4 will be completely transmitted ($T_{14}=1$), while light incident from port 2 or 3 will be completely absorbed ($A_{23}=1-R_{23}-T_{23}=1$). A schematic of the resulting transmission and absorption characteristics is shown in Fig.~\ref{fig:absorb}(a).

To verify these analytical results, we performed numerical simulations of the transmission and reflection spectrum with the rigorous coupled-wave analysis (RCWA) method using a freely available software package \cite{Liu2012}. The structural parameters are identical to the simulation in Fig.~\ref{fig:maxasym}(a), with the difference being the addition of loss in the system. There is a slight shift of the resonance location relative to the Fabry-Perot background due to the different discretization schemes used in the FDTD and RCWA methods. Fig.~\ref{fig:absorb} shows the simulation results. As shown in Fig.~\ref{fig:absorb}(b), when no loss is present (top panel), the transmissivity is close to 1 in the vicinity of the resonance, reaching full transmission $T=1$ at a single point on the Fano resonance as required by $P$-symmetry of the structure (see end of Section 2 for a discussion), and the reflectance is close to 0. With the addition of loss and with waves incident from the port with stronger coupling (incident direction is $\theta=3.5^\circ$ from normal, middle panel), the transmittance is reduced, and at critical coupling the transmittance drops to 0 for the resonance frequency, resulting in full-absorption of the incoming waves. On the other hand, for the same lossy structure and for the opposite incident direction (bottom panel), there is negligible absorption and most of the waves are transmitted. In Fig.~\ref{fig:absorb}(c), we show the maximum absorption for a given incident angle as a function of loss, clearly showing a peak of near-perfect absorption at critical coupling. \textcolor{black}{For other incident angles near that of maximum absorption, the $Q$-matching condition and asymmetric radiation condition are still approximately satisfied, giving rise to high absorption, and the absorption peak will shift to different frequencies following the band dispersion, as shown in \href{link}{Supplement 1}, Fig.~\ref{fig:varyangle}.} Numerically, we find that when light is incident from the port with stronger coupling, the absorption can be as high as $A_{23}=99.8\%$; when light is incident from the port with weaker coupling, the transmission is $T_{14}=99.4\%$ while absorption is only $A_{14}=0.5\%$ (this is roughly equal to the single-pass absorption rate of $0.4\%$ in our simulations). The numerical simulations show excellent agreement with our TCMT predictions for a background Fabry-Perot transmissivity of $t=0.996$ (Fig.~\ref{fig:absorb}(b)); the small difference comes from the contribution of absorption to the Fabry-Perot background. Further numerical optimization placing the resonance frequency closer to the frequency of full transmission on the Fabry-Perot background could further increase on-resonance absorption in the desired port. 

These results are widely applicable to many different absorbing materials. The wide range of achievable resonance quality factors, as discussed in the preceding section, implies that for both strong and weak absorbers, structures can be designed such that there is highly asymmetric coupling and critical coupling, yielding perfect absorption in the system. While our simulations were performed assuming a material with a spatially uniform absorption profile, the generality of the TCMT formalism ensures that it is also applicable to scenarios with only a thin active layer with absorption, such as with 2D materials \cite{Mak2010,Piper2014,Piper2016}.

\section{Discussion and Conclusion}

In conclusion, we developed a temporal coupled-mode theory formalism for general dielectric photonic crystal slab structures with arbitrary in-plane wavevectors, adequately taking into account the time-reversal-symmetry related pair of resonances and coupling channels. Using this formalism, we derived general bounds on the asymmetric radiation rates to the top and bottom of a photonic crystal slab. We then used the intuitions developed from these bounds to show examples of highly asymmetric radiation from inversion-symmetric photonic crystal slabs, demonstrating strong asymmetry, rapid tuning, and a variety of quality factors for different applications. Moreover, we showed how the highly asymmetric coupling to the top and bottom of photonic crystal slabs can be used to achieve perfect absorption for light incident from a single side, for a single-pass absorption rate of less than $0.5\%$, without the need of a back-reflection mirror as in conventional setups. The highly directional radiation could greatly benefit applications such as PCSELs, grating couplers, and LIDARs, while achieving perfect absorption without the need of back-reflection mirrors could increase the efficiency and simplify the design of photodetectors and solar cells. 

\textcolor{black}{While our numerical examples focused on a particular structural design, the general principle of breaking $C_2^z$ symmetry is applicable to a wide range of structures. We now briefly discuss how to implement such structures using readily available fabrication techniques. For example, gratings with slanted walls share the same structural symmetries as those in Fig.~\ref{fig:bound}(a), and thus can approach perfect single-sided radiation and absorption as well. Such gratings can be fabricated using focused ion beam milling \cite{Schrauwen2007}, angled-etching with Faraday cages \cite{Burek2012}, or inclined lithography \cite{DelCampo2008}. More generally, almost any of the techniques for fabricating blazed gratings \cite{Fujita1982,Palmer2005} or constructing 3D photonic crystals \cite{Joannopoulos2011} (e.g. layer-by-layer lithography or holographic lithography) could also be employed in a simplified form to make an asymmetric coating. A wide range of structures breaking $C_2^z$ symmetry and achieving high asymmetry can thus be easily realized with these different techniques.}

Our work provides new design principles for achieving highly directional radiation and perfect absorption in photonics, and could be extended to systems where there are nonlinearities, gain and loss, different substrates and superstrates, and non-reciprocal structures with magneto-optical effects \cite{Wang2005}. Our work can also be generalized to other systems characterized by temporal coupled-mode theory, such as in-plane chiral meta-surfaces, asymmetric ring resonators, and scattering from nano-plasmonic structures.

\section*{Funding and Acknowledgements}
Institute for Soldier Nanotechnologies (W911NF-13-D0001). Solid State Solar Thermal Energy Center (DE-SC0001299). United States-Israel Binational Science Foundation (2013508). National Science Foundation (DMR-1307632). We thank Yong Liang, Ling Lu, Scott Skirlo, Yichen Shen, Aviram Massuda, Emma Regan, Francisco Machado and Nicholas Rivera for helpful discussions. See appendix for supporting material.




\newpage

\begin{appendix}
\section{Detailed Derivation of Temporal Coupled-Mode Theory Equations}

In this section, we present details of the temporal coupled-mode theory (TCMT) formalism in the main text. The TCMT equations are given in the main text as Eq.~(1), and energy conservation and reciprocity considerations constrain the Fabry-Perot direct pathway scattering matrix to take the form given in Eq.~(2) of the main text. In the following discussion, we shall denote the coupling matrices appearing in the main text (constrained by momentum conservation) as
\begin{align}
D&=\begin{pmatrix}
0 & d_1\\d_2 & 0\\0 & d_3\\d_4 & 0\\
\end{pmatrix},\\
K^T&=\begin{pmatrix}
\kappa_1 & 0 & \kappa_3 & 0\\ 0 & \kappa_2 & 0 & \kappa_4\\
\end{pmatrix},
\end{align}
and we shall denote the vector formed by incoming and outgoing waves as $|s_+\rangle$ and $|s_-\rangle$.

First we prove, using reciprocity of the system, that the two resonances related by time-reversal symmetry have the same frequency and decay rate.

For a fixed in-plane momentum $\vec{k}$, the eigenvalue problem with Bloch periodic boundary conditions is given by \cite{Joannopoulos2011}
\begin{align}
\hat{\Theta}_{\vec{k}}\vec{u}_{\vec{k}}(\vec{r})=\frac{\omega(\vec{k})^2}{c^2}\vec{u}_{\vec{k}}(\vec{r}),
\end{align}
where the operator $\hat{\Theta}_{\vec{k}}$ is defined by
\begin{align}
\hat{\Theta}_{\vec{k}}=(i\vec{k}+\nabla)\times\frac{1}{\epsilon(\vec{r})}(i\vec{k}+\nabla)\times,
\end{align}
and $\vec{u}_{\vec{k}}(\vec{r})$ is the Bloch wave function. Due to the coupling to the ports, the modes will have eigenvalues $\omega(\vec{k})$ that are in general complex.

Due to reciprocity, the operators corresponding to the two resonances are related by
\begin{align}
\hat{\Theta}_{\vec{k}}^T=\hat{\Theta}_{\vec{-k}}.
\end{align}
This implies that
\begin{align}
\vec{u}_{\vec{k}}^T\hat{\Theta}_{-\vec{k}}=\frac{\omega_{\vec{k}}^2}{c^2}\vec{u}_{\vec{k}}^T,
\end{align}
so $\vec{u}_{\vec{k}}^T$ is a left eigenvector of the operator $\hat{\Theta}_{-\vec{k}}$. At the same time, we have the eigenvalue problem for $\hat{\Theta}_{-\vec{k}}$ as
\begin{align}
\hat{\Theta}_{-\vec{k}}\vec{u}_{-\vec{k}}=\frac{\omega_{-\vec{k}}^2}{c^2}\vec{u}_{-\vec{k}},
\end{align}
so $\omega_{-\vec{k}}$ and $\omega_{\vec{k}}$ are a pair of left and right eigenvalues for the same operator. However, we know that the left and right eigenvalues of an operator are identical, as can be seen from the characteristic polynomial for any matrix $A$ and $A^T$ being equal:
\begin{align}
\det(A^T-\lambda I)&=\det(A^T-\lambda I^T)=\det((A-\lambda I)^T)\nonumber\\&=\det(A-\lambda I).
\end{align}
We thus conclude that both resonances have the same eigenvalues, i.e. the same frequencies and decay rates.

With this condition, we now prove the constraints on the coupling coefficients mentioned in the main text. We first consider the limit with no absorption loss $\tau_{nr}\rightarrow\infty$. Consider the process where there is no incoming wave $|s_+\rangle=0$ and there is only one resonance that is excited $A_1(0)=A_{10}$, $A_2(0)=0$. By energy conservation, we have
\begin{align}
\frac{d(A_1^* A_1)}{dt}&=-\frac{2}{\tau}A_1^* A_1=-\langle s_-|s_-\rangle=-A_1^* A_1(|d_2|^2+|d_4|^2),
\end{align}
which reduces to (making use of a similar equation for the second resonance)
\begin{align}
|d_2|^2+|d_4|^2=\frac{2}{\tau},|d_1|^2+|d_3|^2=\frac{2}{\tau}.
\end{align}

Now consider the time-reversed process where we send in an exponentially growing wave with initial amplitude $|s_-\rangle^*$ and frequency $\omega_1-\frac{j}{\tau}$. Since the time-reverse of resonance 1 is resonance 2, we expect an exponentially growing amplitude in resonance 2 with no amplitude in resonance 1 or the output ports.

The amplitude equation of resonance 2 reads
\begin{align}\label{eq:tr}
&[j(\omega_1-\frac{j}{\tau})-j\omega_2+\frac{1}{\tau}]A_2=\kappa_2 s'_{2+}+\kappa_4
s'_{4+}\nonumber\\
=&\kappa_2 s_{2-}^*+\kappa_4 s_{4-}^*=\kappa_2 d_2^*A_1^*+\kappa_4 d_4^*A_1^*.
\end{align}

As discussed above, due to reciprocity, the frequencies and decay rates for the two resonances are identical, so using Eq.~(\ref{eq:econs}) and Eq.~(\ref{eq:tr}), we have
\begin{align}\label{eq:kappaeq}
\kappa_2 d_2^*+\kappa_4 d_4^*=d_2d_2^*+d_4d_4^*.
\end{align}

Reciprocity of the system requires that the scattering matrix given by Eq.~(\ref{eq:scatmat}) is symmetric. Since the direct pathway scattering matrix $C$ is symmetric, we must have
\begin{align}
DK^T=KD^T.
\end{align}
The (1,2) entry and (1,4) entry of this equation give
\begin{align}
d_1\kappa_2=d_2\kappa_1,\quad d_1\kappa_4=d_4\kappa_1,
\end{align}
which under the general case where the couplings are nonzero (if they were zero, we could use infinitesimal perturbations to reduce them to the general case) implies
\begin{align}\label{eq:recipkappaeq}
\kappa_2 d_4=\kappa_4 d_2.
\end{align}
Plugging Eq.~(\ref{eq:recipkappaeq}) into Eq.~(\ref{eq:kappaeq}) thus gives us the desired condition that
\begin{align}
\kappa_2=d_2,\quad \kappa_4=d_4,
\end{align}
and by a similar argument $\kappa_1=d_1$,$\kappa_3=d_3$, as quoted in the main text.

We note that the same conclusion can also be obtained following an independent port diagonalization argument presented in Ref. \cite{Suh2004}. There, the resonances and ports are rewritten in normal modes to give independent equations, and the condition that under time-reversal, the ports with no outgoing waves must not couple incoming waves into the resonances imposes conditions similar to Eq.~(\ref{eq:recipkappaeq}).

The condition that there are no outgoing waves in the time-reversed process implies that
\begin{align}
e^{j\phi}\begin{pmatrix}
0 & r & 0 & jt\\r & 0 & jt & 0\\0 & jt & 0 & r\\jt & 0 & r & 0\\
\end{pmatrix}\begin{pmatrix}
0\\d_2^*A_1^*\\0\\d_4^*A_1^*\\
\end{pmatrix}
+\begin{pmatrix}
0&d_1\\d_2&0\\0&d_3\\d_4&0\\
\end{pmatrix}
\begin{pmatrix}
0\\A_1^*
\end{pmatrix}=0,
\end{align}
which expands to two independent equations (the other two equations are linear superpositions of these and their complex conjugates)
\begin{align}
e^{j\phi}(rd_2^*+jtd_4^*)+d_1&=0,\\
e^{j\phi}(jtd_2^*+rd_4^*)+d_3&=0.
\end{align}

We now prove the equivalence between our formalism and the standard multimode TCMT formalism in Ref. \cite{Suh2004} by performing a basis transformation. In the original formalism, the starting point is a Hermitian eigenproblem with time-reversal symmetry, such that the eigenvalues are real:
\begin{align}
\hat{\Theta}\psi=E\psi,
\end{align}
where $\psi$ is a generic wavefunction and $E$ is the energy. Time-reversal symmetry implies that the conjugated wavefunction is also an eigenfunction with the same eigenvalue
\begin{align}
\hat{\Theta}\psi^*=E\psi^*.
\end{align}
Therefore, we can choose symmetric and anti-symmetric combinations of the wavefunctions such that the wavefunction itself is invariant under time-reversal. Under this basis change, any wavefunction definition will be mapped into the standard formalism provided in Ref. \cite{Suh2004}, and the results there shall apply.

More specifically, in our formalism, such a basis choice corresponds to choosing symmetric and anti-symmetric combinations of the resonances with $\vec{k}$ and $-\vec{k}$. We can write down a basis transform matrix into the 
\begin{align}
\textbf{A'}=U\textbf{A},\quad U=\frac{1}{\sqrt{2}}\begin{pmatrix}
1 & 1\\ i & -i
\end{pmatrix},
\end{align}
then the resulting coupling matrices are transformed to be
\begin{align}
K'^T&=UK^T\\ D'&=DU^{-1}=K\sigma_xU^{-1}=K'(U^{-1})^T\sigma_x U^{-1}=K',
\end{align}
which reduces to the same form as Eq.~(5) in Ref. \cite{Suh2004}. This shows that the two formalisms are equivalent up to a basis transformation. In our discussion, we chose the fixed momentum basis so that the physical interpretation of the resonance modes was more transparent.

This basis change is to be contrasted with the formalism in Ref. \cite{Ruan2012}. There, the incoming and outgoing waves are angular momentum channels of a localized nanoparticle, and the formalism is equivalent to the standard treatment up to a basis change of the ports, rather than the resonances. The incorporation of multiple ports on a single side, as described in our paper, could provide new insights into applications derived from two-port analysis for photonic crystal slabs, such as in Ref. \cite{Fan2003,Suh2004a,Wang2013,Piper2014}. In some works \cite{Verslegers2010,Wang2014}, multiple ports have been incorporated, but they correspond to higher diffraction orders, and are often treated as additional loss channels. Our work shows that zeroth order contributions at finite momentum are also important, and could yield interesting new results beyond the standard treatment.

\section{Further Discussion on Transmission and Reflection Spectra}
Generally, the full scattering matrix including the direct pathway and resonance pathway is given by \cite{Suh2004}
\begin{align}\label{eq:scatmat}
S=C+\frac{D K^T}{j(\omega-\omega_0)+\frac{1}{\tau}+\frac{1}{\tau_{nr}}},
\end{align}
here the $\tau$ is labeling the total radiative decay lifetime to all channels of either of the resonances, and $\tau_{nr}$ is the total non-radiative decay lifetime. We have used the fact that the frequencies and decay lifetimes are identical for the two resonances by reciprocity of the system.

The power reflection coefficient corresponds to the amplitude squared of the (1,2) element of the scattering matrix. Using the fact that $\kappa_i=d_i$, we have
\begin{align}\label{eq:R}
R=|S_{12}|^2=\left|e^{j\phi}r+\frac{d_1d_2}{j(\omega-\omega_0)+\frac{1}{\tau}+\frac{1}{\tau_{nr}}}\right|^2.
\end{align}
This expression can be simplified to a form that explicitly exhibits features such as full transmission and/or full reflection, and eliminates explicit dependence on the phases of the coupling rates. This is accomplished by taking the norm of the time-reversal relations.

First, let us write the decay rates to different ports as $d_i=\sqrt{\frac{2}{\tau_i}}e^{j\theta_i}$. We wish to find $\theta_1+\theta_2$, which will allow us to evaluate the preceding expression. Eqs.~(\ref{eq:fulld1}, \ref{eq:fulld2}) can be written as
\begin{align}
r\sqrt{\frac{2}{\tau_2}}e^{-j\theta_2}+jt\sqrt{\frac{2}{\tau_4}}e^{-j\theta_4}+\sqrt{\frac{2}{\tau_1}}e^{j\theta_1}&=0,\\
jt\sqrt{\frac{2}{\tau_2}}e^{-j\theta_2}+r\sqrt{\frac{2}{\tau_4}}e^{-j\theta_4}+\sqrt{\frac{2}{\tau_3}}e^{j\theta_3}&=0.
\end{align}

These equations can be used to prove that $|d_1|^2+|d_3|^2=|d_2|^2+|d_4|^2$. This, combined with Eq.~(\ref{eq:econs}), implies self-consistently that the two resonances have the same decay rate.

Defining $\alpha=\theta_1+\theta_2$, $\beta=\theta_1+\theta_4$, the preceding equation reads
\begin{align}\label{eq:alpha}
\sqrt{\frac{2}{\tau_1}}&=-r\sqrt{\frac{2}{\tau_2}}e^{-j\alpha}-jt\sqrt{\frac{2}{\tau_4}}e^{-j\beta},\\
\sqrt{\frac{2}{\tau_3}}e^{j\theta_3-j\theta_1}&=-jt\sqrt{\frac{2}{\tau_2}}e^{-j\alpha}-r\sqrt{\frac{2}{\tau_4}}e^{-j\beta}.
\end{align}

Taking the norm squared of both sides, we obtain
\begin{align}
\frac{2}{\tau_1}&=\frac{2r^2}{\tau_2}+\frac{2t^2}{\tau_4}-\frac{4rt}{\sqrt{\tau_2\tau_4}}\sin(\alpha-\beta)\\
\frac{2}{\tau_3}&=\frac{2r^2}{\tau_4}+\frac{2t^2}{\tau_2}+\frac{4rt}{\sqrt{\tau_2\tau_4}}\sin(\alpha-\beta)
\end{align}

Adding these two equations, we find
\begin{align}
\frac{1}{T_2}=\frac{1}{\tau_1}+\frac{1}{\tau_3}=\frac{1}{\tau_2}+\frac{1}{\tau_4}=\frac{1}{T_1},
\end{align}
as desired. 

Now we solve for the transmission spectrum by solving $\alpha$. Moving the term containing $\alpha$ over to the left hand side in Eq.~(\ref{eq:alpha}) and taking the norm squared gives
\begin{align}
\frac{2t^2}{\tau_4}=\frac{2r^2}{\tau_2}+\frac{2}{\tau_1}+\frac{4r\cos\alpha}{\sqrt{\tau_1\tau_2}}.
\end{align}

To simplify the preceding expression, we write 
\begin{align}\label{eq:deftau}
\frac{1}{\tau}=\frac{1}{\tau_1}+\frac{1}{\tau_3}=\frac{1}{\tau_2}+\frac{1}{\tau_4},\quad\frac{1}{\sigma}=\frac{1}{\tau_1}-\frac{1}{\tau_4},
\end{align}
and thus
\begin{align}\label{eq:defcos}
\cos\alpha&=\frac{\sqrt{\tau_1\tau_2}}{2r}\left( -\frac{r^2}{\tau}-\frac{1}{\sigma}\right)\\
\sin\alpha&=\pm\sqrt{1-\frac{\tau_1\tau_2}{4r^2}\left( \frac{r^2}{\tau}+\frac{1}{\sigma}\right)^2}\nonumber\\&=\pm\frac{\sqrt{\tau_1\tau_2}}{2}\sqrt{\frac{4}{\tau_1\tau_2}-\frac{r^2}{\tau^2}-\frac{2}{\tau\sigma}-\frac{1}{\sigma^2 r^2}}\label{eq:defsin}
\end{align}

The reflectance given by Eq.~(\ref{eq:R}) can be written as
\begin{align}\label{eq:Rres}
R&=\frac{|rj(\omega-\omega_0)+\frac{r}{\tau}+\frac{2}{\sqrt{\tau_1\tau_2}}(\cos\alpha+j\sin\alpha)|^2}{(\omega-\omega_0)^2+\frac{1}{\tau^2}}\nonumber\\
&=\frac{r^2(\omega-\omega_0)^2+\frac{r^2}{\tau^2}+\frac{4}{\tau_1\tau_2}+\frac{4r\cos\alpha}{\tau\sqrt{\tau_1\tau_2}}+\frac{4r(\omega-\omega_0)\sin\alpha}{\sqrt{\tau_1\tau_2}}}{(\omega-\omega_0)^2+\frac{1}{\tau^2}}\nonumber\\
&=\frac{\left[r(\omega-\omega_0)\pm\sqrt{\frac{4}{\tau_1\tau_2}-\frac{r^2}{\tau^2}-\frac{2}{\tau\sigma}-\frac{1}{\sigma^2 r^2}}\right]^2+(\frac{1}{\sigma r})^2}{(\omega-\omega_0)^2+\frac{1}{\tau^2}},
\end{align}
where the last line can be directly verified by expanding all terms using Eqs.~(\ref{eq:deftau}-\ref{eq:defsin}) given above. This expression is similar to Eq.~(12) of Ref. \cite{Wang2013}, except that here because $\tau_1$ and $\tau_2$ are not necessarily identical, the form of the expression within the square root is different.

The expression in the square root in Eq.~(\ref{eq:Rres}) is given by
\begin{align}
&\frac{4}{\tau_1\tau_2}-\frac{2}{\tau\sigma}-\frac{r^2}{\tau^2}-\frac{1}{\sigma^2r^2}\nonumber\\
&=\frac{4}{\tau_1\tau_2}-2\left(\frac{1}{\tau_1}-\frac{1}{\tau_4}\right)\left(\frac{1}{\tau_2}+\frac{1}{\tau_4}\right)-\frac{r^2}{\tau^2}-\frac{1}{\sigma^2r^2}\nonumber\\
&=\frac{2}{\tau_1\tau_2}+\frac{2}{\tau_4^2}+\frac{2}{\tau_4}\left(\frac{1}{\tau_2}-\frac{1}{\tau_1}\right)-\frac{r^2}{\tau^2}-\frac{1}{\sigma^2r^2},
\end{align}
which agrees with Ref. \cite{Wang2013} when the system possesses $C_2^z$ symmetry such that $\tau_1=\tau_2$, $\tau_3=\tau_4$.

Let us now consider the consequences of this expression on frequencies of full reflection and full transmission.

For full transmission, we require $T=1$ and thus $R=0$. Comparing to Eq.~(\ref{eq:R}) shows that
\begin{align}
\frac{1}{\sigma}&=\frac{1}{\tau_1}-\frac{1}{\tau_4}=0\Rightarrow\tau_1=\tau_4,\\
\omega-\omega_0&=\pm\frac{1}{r}\sqrt{\frac{4}{\tau_4\tau_2}-\left(\frac{1}{\tau_2}+\frac{1}{\tau_4}\right)^2r^2},
\end{align}
so full transmission occurs when the decay rates are effectively $P$-symmetric. For a structure that possesses $P$-symmetry, this condition is automatically guaranteed, and indeed we observe that transmission reaches unity near the Fano resonance, see Fig. 1(b) of the main text.

For full reflection, we require $R=1$. Eq.~(\ref{eq:Rres}) imposes the condition
\begin{align}
&\left[r(\omega-\omega_0)\pm\sqrt{\frac{4}{\tau_1\tau_2}-\frac{r^2}{\tau^2}-\frac{2}{\tau\sigma}-\frac{1}{\sigma^2r^2}}\right]^2+\left(\frac{1}{\sigma r}\right)^2\nonumber\\&=(\omega-\omega_0)^2+\frac{1}{\tau^2}.
\end{align}

The discriminant of this equation is
\begin{align}
&4\left(\frac{4}{\tau_1\tau_2}-\frac{r^2}{\tau^2}-\frac{2}{\tau\sigma}-\frac{1}{\sigma^2 r^2}\right)-4(1-r^2)\left(\frac{1}{\tau^2}-\frac{1}{\sigma^2 r^2}\right)\nonumber\\
=&-4\left(\frac{1}{\tau_1}-\frac{1}{\tau_2}\right)^2,
\end{align}
so full reflection is possible when the discriminant is non-negative, i.e. when the decay rates are effectively $C_2^z$-symmetric: $\tau_1=\tau_2$. For a structure that possesses $C_2^z$-symmetry, this condition is guaranteed, as discussed in Ref. \cite{Wang2013}.

\begin{figure}[htb]
\centering
\includegraphics[width=\linewidth]{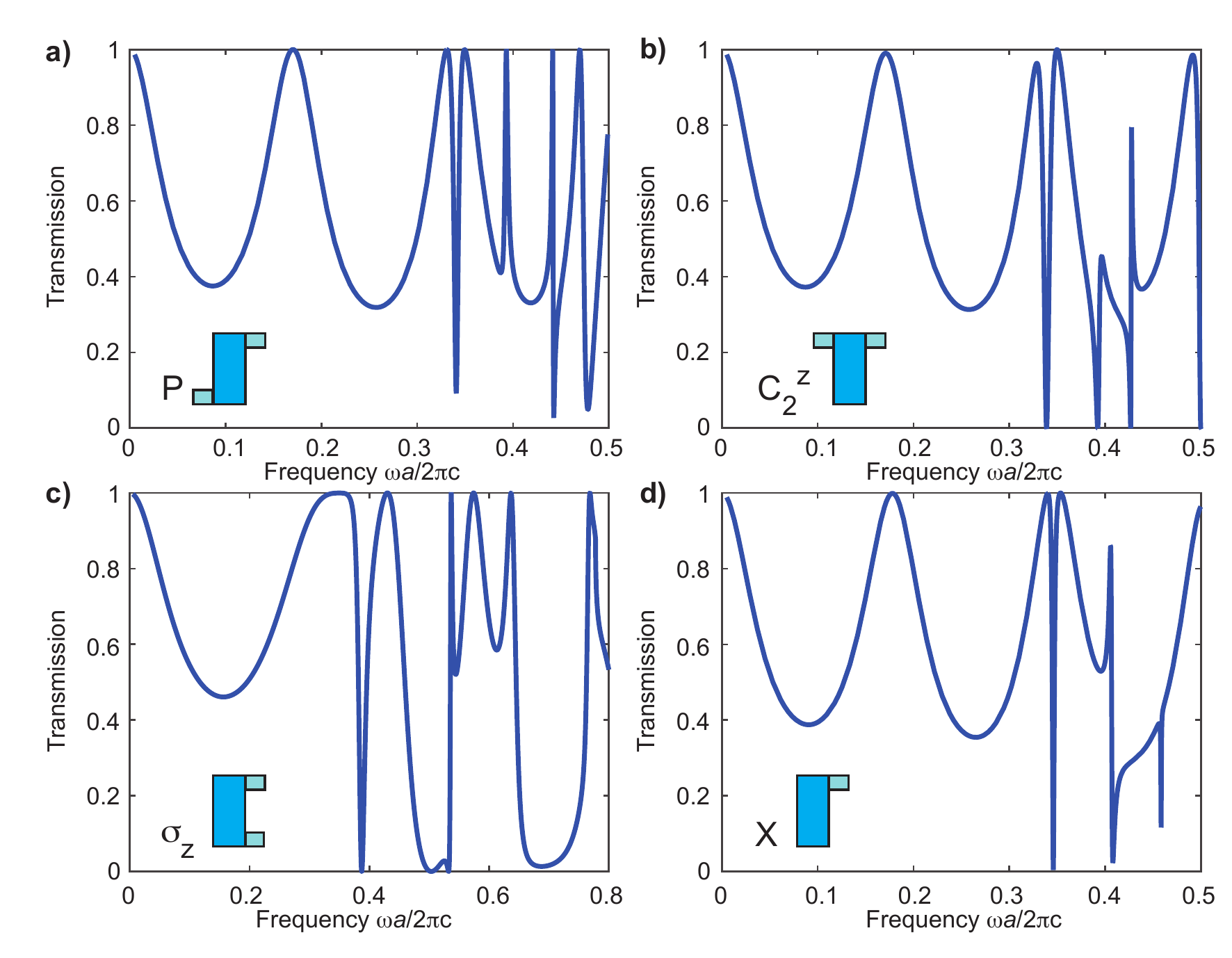}
\caption{Transmission spectra for different structure symmetries. (a) Inversion-symmetric structure, showing points of full transmission on the Fano resonance; (b) $C_2^z$-symmetric structure, showing points of full reflection on the Fano resonance; (c) $z$-mirror-symmetric structure, showing points of full transmission and full reflection on the Fano resonance; (d) Structure without geometric symmetries, showing no points of full transmission or reflection for generic resonances. \label{fig:multispec}}
\end{figure}

We show examples of transmission spectra in Fig.~\ref{fig:multispec}, for different symmetries of the underlying structure, verifying our preceding discussion. Note that the structure does not have to possess the corresponding symmetry for the coupling rates to possess the symmetries. As an example, consider a structure that has $\sigma_z$ symmetry, i.e. it is mirror-symmetric in the $z$-direction (Fig.~\ref{fig:multispec}(c)). $\sigma_z$-symmetry imposes $\tau_1=\tau_3$, $\tau_2=\tau_4$; reciprocity then requires that $1/\tau_1+1/\tau_3=1/\tau_2+1/\tau_4$, so that $\tau_1=\tau_2=\tau_3=\tau_4$, and thus even though the structure itself does not possess $P$ or $C_2^z$ symmetry, the coupling rates do, so the transmission spectrum should have Fano resonances with points of full transmission/reflection. This is indeed what we observe in Fig.~\ref{fig:multispec}(c). On the other hand, for more general structures, most resonances will not have such points of full transmission/reflection, as shown in Fig.~\ref{fig:multispec}(d).

We now discuss the transmission spectrum of an inversion-symmetric structure in the presence of loss. We assume that the background amplitude transmission coefficient $t$ is close to 1, and that the coupling coefficients saturate the bound $|d_2|/|d_1|=\sqrt{(1+t)/(1-t)}$. As $|d_1|^2+|d_2|^2=2/\tau$, we have that
\begin{align}
d_1=\sqrt{\frac{1-t}{\tau}}e^{j\alpha}, d_2=\sqrt{\frac{1+t}{\tau}}e^{j\beta},
\end{align}
where $\alpha$ and $\beta$ characterize the complex phase of the coupling coefficient. Plugging this into the time-reversal symmetry constraints Eqs.~(4, 5) of the main text, we find that
\begin{align}
(1-t)e^{-j\alpha}+jte^{-j\beta}+e^{j\beta}&=0,\\
(1+t)e^{-j\beta}+jte^{-j\alpha}+e^{j\alpha}&=0.
\end{align}
Eliminating $\alpha$ using one equation and plugging into the other, we find that $\beta=-\pi/4$ and $\alpha=-3\pi/4$.

On resonance and with the radiative quality factor and non-radiative quality factor matched, we have
\begin{align}
R_{14}=R_{23}&=|r+\frac{re^{j(\alpha+\beta)}/\tau}{2/\tau}|^2=\frac{r^2}{4},\\
T_{14}&=|jt+\frac{(1-t)e^{2j\alpha}/\tau}{2/\tau}|^2=(\frac{1+t}{2})^2,\\
T_{23}&=|jt+\frac{(1+t)e^{2j\beta}/\tau}{2/\tau}|^2=(\frac{1-t}{2})^2.
\end{align}
The absorption can be calculated by subtracting transmission and reflection from unity, giving rise to Eqs.~(12,13) of the main text.

\section{Further Discussion on Theoretical Bounds}

The general expression for the bound without any assumptions of the symmetry of the structure is given by Eq.~(10) of the main text.

If we make the assumption of $C_2^z$ symmetry, the two channels on the top and bottom would be constrained to have the same coupling rates, and hence $d_1=d_2$, $d_3=d_4$. Plugging this into Eq.~(4), we have
\begin{align}
e^{j\phi}(rd_1^*+jtd_3^*)+d_1=0,
\end{align}
which reduces to the same bound as in Ref. \cite{Wang2013}:
\begin{align}\label{eq:wangbound}
\sqrt{\frac{1-r}{1+r}}\leq a_l=a_r\leq\sqrt{\frac{1+r}{1-r}},
\end{align}
thus demonstrating consistency of our approach. This result can also be obtained from Eq.~(10) by setting $a_l=a_r$.

\begin{figure}[htb]
\begin{center}
\includegraphics[width=\linewidth]{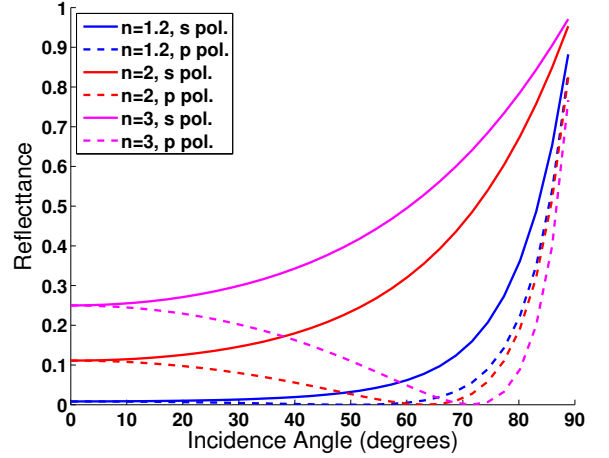}
\caption[Reflectance calculated from Fresnel equations]{Reflectance calculated according to Fresnel equations for light incident from air ($n_0=1$) to dielectric.\label{fig:fresnel}}
\end{center}
\end{figure}

Fig.~\ref{fig:fresnel} shows a calculation of the interface reflection coefficient based on Fresnel equations. The resulting Fabry-Perot transmittance is given by \cite{Hecht2001}
\begin{align}
t^2=\frac{1}{1+\frac{4R_0\sin^2(\delta/2)}{(1-R_0)^2}},
\end{align}
where $R_0$ is the interface reflectance and $\delta$ is the phase accumulated when reflecting between the interfaces. It is evident that the reflectance will generally be small for realistic dielectric contrasts at optical wavelengths, so the bound given in Eq.~(\ref{eq:wangbound}) will impose relatively strict constraints.

However, when the structure does not possess $C_2^z$ symmetry, Eq.~(\ref{eq:wangbound}) no longer applies. As an example, when the structure is inversion-symmetric, the decay rates must satisfy $d_1=d_4$, $d_2=d_3$, $a_l=1/a_r$, which would give
\begin{align}
\frac{\tau_3}{\tau_1}=|\frac{d_1}{d_3}|^2=|\frac{-re^{j\phi-j\theta_2}}{e^{j\theta_1}+jte^{j\phi-j\theta_4}}|^2=\frac{1-t^2}{(1+t\cos\phi')^2},
\end{align}
where $\phi'$ is a phase that in general can be tuned through an entire $2\pi$ range, and we have defined $\tau_i=|d_i|^2$ to characterize the loss rate. The asymmetry is thus bounded by 
\begin{align}
\sqrt{\frac{1-t}{1+t}}\leq a_l=\frac{1}{a_r}\leq \sqrt{\frac{1+t}{1-t}},
\end{align}
where again $t$ is the amplitude transmission coefficient of the direct process governed by the Fabry-Perot cavity response. As discussed in the main text, this will allow arbitrarily asymmetric radiation.

\begin{figure}[htb]
\begin{center}
\includegraphics[width=\linewidth]{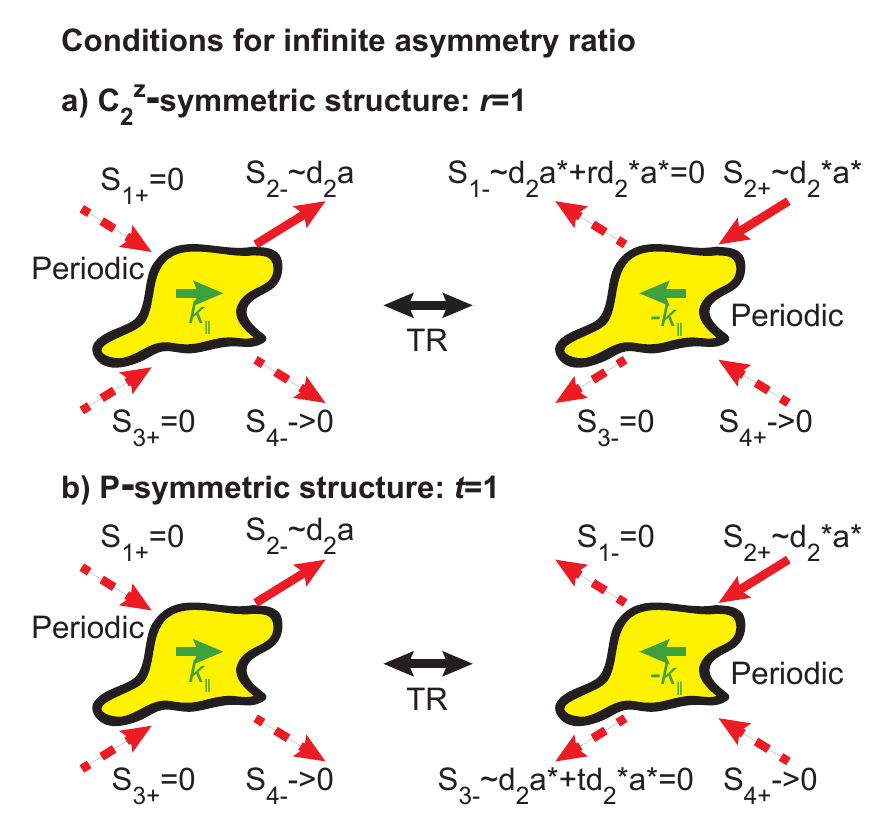}
\caption{Conditions for achieving infinite asymmetry in (a) $C_2^z$-symmetric and (b) $P$-symmetric structures.\label{fig:infinasym}}
\end{center}
\end{figure}

The intuition for the theoretical bounds Eqs.~(\ref{eq:wangbound}, \ref{eq:tbound}) is shown in Fig.~\ref{fig:infinasym}. Depicted here is the condition for achieving an infinite asymmetry ratio. This can also be viewed as a single-side bound state in the continuum \cite{Hsu2013a,Hsu2013b,Zhen2014}, in which the resonance exclusively radiates to one side of the slab and not the other. Consider the scenario where the resonance at $\vec{k}_\parallel$ is excited, and the resonance decays to ports 2 and 4. Since the asymmetric coupling ratio is infinity, all of the power will decay into port 2. After time-reversal, the condition that there are no outgoing waves requires that the direct transmission/reflection for incident waves from port 2 cancels the decay from the resonance at $-\vec{k}_\parallel$. As we can see from Fig.~\ref{fig:infinasym}, this would require $r=1$ for a $C_2^z$-symmetric structure, and $t=1$ for a $P$-symmetric structure. Again, we note that for realistic material refractive indices, the latter is much more readily achievable.

For some practical devices such as lasers, we cannot neglect the $\vec{k}_\parallel'=-\vec{k}_\parallel$ vector that has an opposite momentum, and should consider the total asymmetry ratio after taking both sides into account. Using the expression $|d_1|^2+|d_3|^2=|d_2|^2+|d_4|^2$ resulting from the time-reversal-pair relation between the resonances, the total asymmetric ratio can be written as
\begin{align}\label{eq:totalasym}
a^2=\frac{|d_3|^2+|d_4|^2}{|d_1|^2+|d_2|^2}=\frac{a_r^2+a_l^2+2a_l^2a_r^2}{a_l^2+a_r^2+2},
\end{align}
where $a_l$ and $a_r$ satisfy the bound specified in Eq.~(10). Taking the partial derivative of the above equation with respect to $a_l^2$ yields
\begin{align}
\frac{\partial}{\partial(a_l^2)}(\frac{a_r^2+a_l^2+2a_r^2a_l^2}{a_r^2+a_l^2+2})
=\frac{2(a_r^2+1)^2}{(a_l^2+a_r^2+2)^2}>0,
\end{align}
so to maximize the asymmetric ratio requires minimizing $a_l$ within the bounds. We thus plug in the lower bound of Eq.~(10) into Eq.~(\ref{eq:totalasym}), which gives
\begin{align}
a^2&=\frac{a_r^2+(\frac{t+ra_r}{r-ta_r})^2+2}{a_r^2+(\frac{t+ra_r}{r-ta_r})^2+2a_r^2(\frac{t+ra_r}{r-ta_r})^2}\nonumber\\
&=\frac{(a_r^2+2)(r-ta_r)^2+(t+ra_r)^2}{a_r^2(r-ta_r)^2+(2a_r^2+1)(t+ra_r)^2}.
\end{align}

Taking the derivative of the above expression with respect to $a_r$, we find
\begin{align}
\frac{d(a^2)}{d(a_r^2)}&=\frac{-8a_r+8t^2a_r+4tr(a_r^2-1)}{(2tra_r+2a_r^2-t^2(a_r^2-1))^2},
\end{align}
which reaches extrema at $a_r^2=(r\pm 1)/t$. Retaining the physical solution (+ sign), the maximal amount of asymmetry is attained when $a_l^2=a_r^2=(r+1)/t$. This suggests that the maximal amount of top-down asymmetry still observes the bound presented in Eq.~(\ref{eq:wangbound}).


Our derivation relies on the assumption of time-reversal symmetry and energy conservation, so one possibility to go beyond this bound of total coupling is to relax these assumptions and generalize TCMT to cases with magneto-optical effects and/or gain and loss, as well as cases where the substrate and superstrate are different. A more practical way to get around this might be to employ side-coupling or laser pumping, in which the incident light imposes a certain $\vec{k}_\parallel$ to the system.

\section{Simulation Details}

\begin{figure}[htb]
\begin{center}
\includegraphics[width=\linewidth]{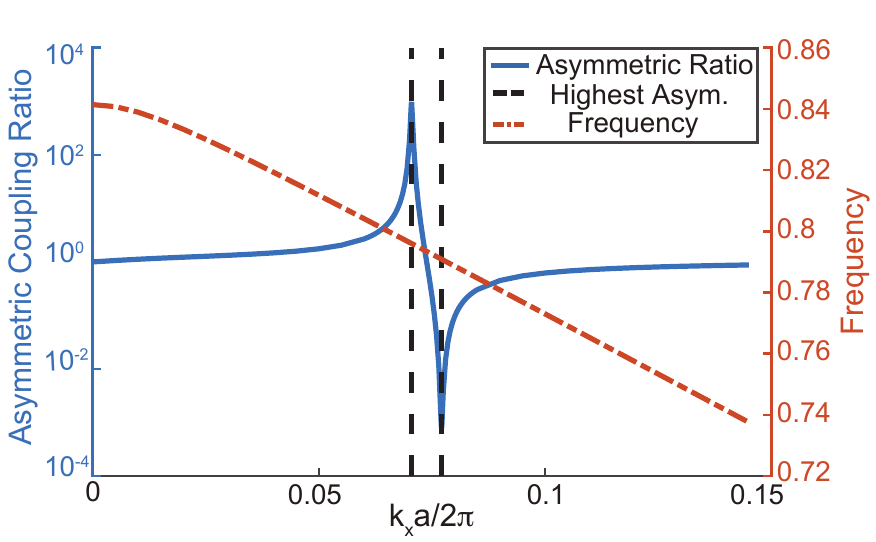}
\caption{Frequency spectrum and asymmetry ratio for the same structure as Fig. 3(b) of the main text.\label{fig:freq}}
\end{center}
\end{figure}

To verify these analytical results, we performed numerical simulations using the finite difference time domain (FDTD) method \cite{Taflove2013} with the freely available software package MEEP \cite{Oskooi2010}. Harmonic inversion \cite{Mandelshtam1997} and far field amplitude analysis are used to determine the coupling rates to top and bottom for each resonance. We place reference planes in the far field, between the structure and the perfectly matched layers (absorbs outgoing waves), and integrate the Bloch wave functions over a unit cell on this plane, as described in Ref. \cite{Zhen2014}. This gives a complex vector $(c_x,c_y)$ containing amplitude and phase information of the coupling rates. We examine the amplitude to determine the asymmetry of the coupling rates, but the phase variation across different $\vec{k}_\parallel$-vector points also contains interesting information in connection to bound states in the continuum.

To compare to the bound given by Eq.~(\ref{eq:tbound}), we use plane wave excitation to determine the transmission and reflection spectrum for different frequencies and wavevectors, and fit the Fabry-Perot background away from the resonance to determine the background transmittivity. Due to the differences between discretization schemes in various software packages, we used MEEP for both resonance calculations and transmission spectrum calculations. Convergence for transmission spectra near resonances is relatively poor in MEEP, so we use narrow-band excitations of width $\Delta\omega=0.01\times 2\pi c/a$ to excite plane waves and discard data close to the resonance. We vary the refractive index, height and width of the dielectric and optimize the maximal asymmetric coupling with respect to wavevector for each structure; see Fig. 2(a) in the main text for a schematic of our simulation parameters.

To verify the TCMT bounds, we choose a set of structures with $a=1$, $h=1.5$, $w=0.45$, $n_0=1.45$, and vary the values of $n_d$ and $d$. For each set of structural parameters, we optimize over all $\vec{k}_\parallel$ to find the direction of largest asymmetric coupling and extract the background transmission coefficient, which we then use to verify the TCMT bounds. A typical frequency spectrum for the second TE band of the photonic crystal we consider is shown in Fig.~\ref{fig:freq}.

\begin{figure}[htb]
\begin{center}
\includegraphics[width=\linewidth]{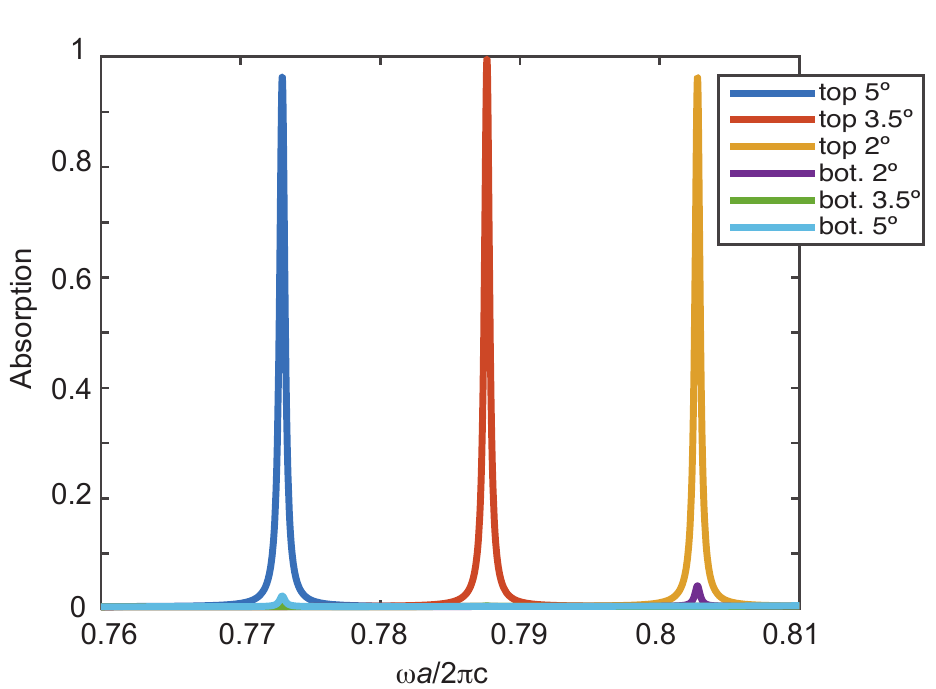}
\caption{Absorption spectrum for the same structure as Fig.~4 in the main text, for different incident angles around the incident direction of full absorption/transmission.\label{fig:varyangle}}
\end{center}
\end{figure}

To verify the possibility of achieving highly efficient absorption based on a structure with highly directional radiation, we performed simulations using the rigorous coupled-wave analysis (RCWA) method with a freely available software package \cite{Liu2012}. The RCWA method allows faster computation of the spectrum and is more robust near the resonances, but the discretization is performed by Fourier space truncation instead of real-space averaging, giving rise to slightly different results than FDTD simulations. However, using the same parameters as above of an example with highly asymmetric radiation, we find that the background reflecttance is not shifted too much, and highly efficient absorption can be achieved based on this design, see the discussion and Fig.~4 in the main text. In Fig.~\ref{fig:varyangle}, we show the absorption spectrum for the same structure as Fig.~4 in the main text. When the change of the incident angle is small, the Q-matching condition and the asymmetric radiation condition for the resonance are still approximately satisfied, leading to high absorption or transmission near the resonance. Furthermore, the peak of absorption will shift to different frequencies under different incident angles, following the band dispersion.

\section{Experimental Considerations and Perspectives}
In the preceding sections, we have derived a TCMT formalism to understand the coupling of photonic crystal slabs to arbitrary in-plane $\vec{k}$-vectors, and used this formalism to derive a bound on the achievable asymmetric coupling from photonic crystal slabs. Examples of strong asymmetry have also been given guided by these design principles.

Fabrication of photonic crystal slabs often uses electron beam lithography with reactive ion etching, interference lithography \cite{Lee2014}, or nanoimprint technology \cite{Chou1996}. These methods generally approach the sample from above, resulting in straight vertical walls in the fabricated devices. For our proposed devices that break $z$-mirror symmetry while preserving inversion symmetry, other fabrication methods will be required. One possibility would be to use focused ion beam milling to produced slanted walls \cite{Schrauwen2007}. Recently developed technologies such as angled-etching \cite{Burek2012} are also a feasible alternative.


For laser applications \cite{Hirose2014}, it is important to address the symmetry between $\vec{k}_\parallel$ and $-\vec{k}_\parallel$. In a system with inversion symmetry, the emission to the top at $\vec{k}_\parallel$ and to the bottom at $-\vec{k}_\parallel$ will be of equal magnitude, which implies that although for a fixed $\vec{k}$-vector we have achieved strong asymmetry, for an actual laser we still need some mechanism to separate the pairs of $\vec{k}$-vectors related by inversion/time-reversal symmetry. One possibility is to operate in an amplifier mode, in which an incoming seed laser stimulates laser emission; due to the properties of stimulated emission, the photons produced must be identical to the incoming ones and thus have the same $\vec{k}_\parallel$, thereby separating the pairs of modes. Another possibility is to use a tilted pump profile (either on-resonance optical pumping or electrical gating) that breaks inversion symmetry, thus enabling preferential lasing of one mode.

Another important question that will need to be addressed is control of the $\vec{k}$-point of lasing and achieving strong asymmetry at this particular point. Generally, lasing will occur in the direction of highest total quality factor. We have demonstrated that asymmetric coupling ratios as large as 300 can be achieved at the point of highest quality factor, but the quality factor profile is relatively flat around the neighborhood. An alternative way to achieve lasing at the desired $\vec{k}$-point is to use other dissipation channels to control the total quality factor. The transverse quality factor resulting from the finite boundary area of the photonic structure may often provide a larger contribution to the total decay, and by engineering the group velocity by band folding techniques \cite{Kurosaka2010}, it may be possible to move the lasing mode to a point of stronger asymmetry.

For grating couplers, other factors such as mode matching with the optical fiber that light will be coupled to, as well as reflection back into the waveguide, both need to be taken care of. Mode matching can be achieved by chirping the lattice parameters such that the quality factor further along the waveguide is smaller, thereby emitting the same total amount of light; reflections are already partially mitigated by tilted walls, and further impedance matching could help to reduce such loss. However, more extensive numerical optimization and FDTD simulations will be required, and for CMOS-compatible platforms the effect of substrates also needs to be incorporated into our formalism.

While the bounds corresponding to $(1+t)/(1-t)$ have been derived under the assumption of an inversion-symmetric structure, it might also be possible to engineer such strong asymmetries in a structure that does not retain this symmetry. It will be interesting to explore more deeply the possibilities for achievable asymmetry in such a more general setting and demonstrate strong asymmetry for more general device configurations.
\end{appendix}

\end{document}